\documentclass[useAMS,usenatbib]{mn2e}

\usepackage{graphicx}

\title[Warped accretion disc in NGC\,4258]{Is the Bardeen-Petterson effect responsible for the warping and precession in NGC\,4258?}
\author[A. Caproni, Z. Abraham, M. Livio \& H. J. Mosquera Cuesta]{A. Caproni$^{1,2,3}$\thanks{E-mail:
anderson.caproni@unicsul.br}, Z. Abraham$^{2}$, M. Livio$^{3}$ \& H. J. Mosquera Cuesta$^{4,5}$\\
$^{1}$N\'ucleo de Astrof\'\i sica Te\'orica, CETEC, Universidade Cruzeiro do Sul, R. Galv\~ao Bueno 868, 01506-000, S\~ao Paulo, SP, Brazil\\
$^{2}$Instituto de Astronomia, Geof\'\i sica e Ci\^encias Atmosf\'ericas, Universidade de S\~ao Paulo, R. do Mat\~ao 1226, \\ Cidade Universit\'aria, 
CEP 05508-900, S\~ao Paulo, SP, Brazil\\
$^{3}$Space Telescope Science Institute, 3700 San Martin Drive, Baltimore, MD, USA, 21218\\
$^{4}$Instituto de Cosmologia, Relatividade e Astrof\'{\i}sica (ICRA-BR),  Centro Brasileiro de Pesquisas F\'\i sicas, 
R. Dr. Xavier Sigaud 150,\\ 22290-180, Rio de Janeiro, RJ, Brazil\\
$^{5}$Abdus Salam International Centre for Theoretical Physics, Strada Costiera 11, Miramare 34014, Trieste, Italy}
\begin{document}

\date{}

\pagerange{\pageref{firstpage}--\pageref{lastpage}} \pubyear{2002}

\maketitle

\label{firstpage}

\begin{abstract}
   Strong evidence for the presence of a warped Keplerian accretion disc in 
   NGC\,4258 (M\,106) has been inferred from the kinematics of water masers 
   detected at sub-parsec scales. Assuming a power-law accretion disc and 
   using constraints on the disc parameters derived from observational data, 
   we have analyzed the relativistic Bardeen-Petterson effect driven by a 
   Kerr black hole as the potential physical mechanism responsible for the 
   disc warping. We found that the Bardeen-Petterson radius is comparable 
   to or smaller than the inner radius of the maser disc (independent of the 
   allowed value for the black hole spin parameter). 
   Numerical simulations for a wide range of physical conditions have shown 
   that the evolution of a misaligned disc 
   due to the Bardeen-Petterson torques usually produces an inner flat disc 
   and a warped transition region with a smooth gradient in the tilt and twist 
   angles. Since this structure is similar to that seen in NGC\,4258, 
   we propose that the Bardeen-Petterson effect may be responsible for the 
   disc warping in this galaxy. 
   We estimated 
   the time-scale necessary for the disc inside of the Bardeen-Petterson 
   radius to align with the black hole's equator, as a function of the 
   black hole spin. Our results show that the Bardeen-Petterson effect 
   can align the disc within a few billion years in the case of NGC\,4258. 
   Finally, we show that if the observed curvature of the outer anomalous 
   arms in the galactic disc of NGC\,4258 is associated with the precession 
   of its radio jet/counterjet, then the Bardeen-Petterson effect can provide 
   the required precession period.
\end{abstract}


\section{Introduction}

   The barred spiral galaxy NGC\,4258, located at a distance of 7.2$\,\pm\,$0.3 Mpc 
   \citep{her99}, presents a well-studied active nuclear region, classified as a 
   Seyfert 1.9, according to \citet{ho97}, with a warped sub-parsec 
   Keplerian water maser accretion disc surrounding a supermassive black hole 
   \citep{miy95,her05}. 
   \citet{nema95} modelled the maser region as a warped viscous accretion disc 
   that is illuminated obliquely by a central X-ray source, so that the disc 
   heating is dominated by X-ray photons. They also suggested that the edges of 
   the water maser region are determined by the thermodynamical conditions of the disc 
   (and this has been corroborated recently by \citealt{her05}). 
   Low-energy X-rays from the galactic centre are absorbed by a hydrogen column 
   density of approximately $10^{23}$ cm$^{-2}$, 
   providing indirect evidence that the warped accretion disc is the X-ray 
   absorber in NGC\,4258 \citep{fru05}. 

   High-resolution interferometric images at radio wavelengths show the presence 
   of a compact jet \citep{her97}, and of two radio spots located at 840 pc south 
   and 1.7 kpc north from the nucleus and identified as bow shock counterparts of 
   the jet-ambient interaction \citep{cec00}.

   The maser disc, which extends roughly between 0.14 and 0.28 pc in relation to 
   the nucleus \citep{her05}, is counterrotating with respect to the galactic disc 
   of the host galaxy, in which kilo-parsec anomalous arms have been detected at 
   the H$\alpha$ transition \citep{cocr61} and later at radio to X-ray frequencies
   (e.g., \citealt{bur63,kru72,alhu82,cec00,wil01}).

   A detailed study of the kinematics of the maser spots conducted by \citet{her05} 
   showed a deviation from Keplerian motion in their projected rotation curve of 
   about 9 km s$^{-1}$. This deviation was successfully modeled as a Keplerian 
   warped accretion disc, with a radial gradient in its inclination of 
   approximately 0.034 mas$^{-1}$.

   Although the existence of a warped accretion disc in NGC\,4258 seems to be 
   the best explanation for the observational data, the nature of the physical 
   mechanism responsible for producing such a structure has not been unambiguously 
   established. 
   
   Tidal torques in a massive binary system were proposed by 
   \citet{pap98} as being responsible for the disc warping (in their model, the 
   companion object has a mass comparable to that of the disc). 
   
   In addition, 
   irradiation-driven torques produced by non-axisymmetric forces due 
   to radiation pressure \citep{prin96} were also invoked to explain the 
   observed behaviour. As noted by \citet{mal96} and \citet{mal98}, if the 
   accretion disc of NGC\,4258 is radiatively inefficient \citep{las96,gam99}, 
   this mechanism cannot produce the inferred warp in NGC\,4258, since 
   the disc becomes radiatively unstable far from its maser region. If 
   the disc is radiatively efficient but the shear acting vertically 
   does not have the same value as the azimuthal shear, the critical radius 
   beyond which the radiative instability acts will increase. For a 
   viscosity parameter $\alpha\la 0.2$ \citep{shsu73}, the masing disc of 
   NGC\,4258 is stable against the radiation instability \citep{gam99}. 
   
   In this work, we analyze an alternative mechanism that could warp the 
   accretion disc of NGC\,4258: the Bardeen-Petterson effect \citep{bape75}, 
   predicted in the framework of general relativity when the spin 
   axis of a Kerr black hole is inclined in relation to the angular momentum 
   vector of the disc. A brief description of the effect is presented in 
   Section 2. In Section 3, we present our accretion disc model, introducing 
   its basic parameters, as well as their values constrained by the 
   observational data of NGC\,4258. The Bardeen-Petterson radius and the 
   alignment time-scale between the accretion disc and the spinning black 
   hole, calculated in terms of the black hole spin and for different disc 
   parameters, are presented in Section 4. The possible connection between 
   the radio jet of NGC\,4258 and the anomalous arms in the galactic disc 
   is also discussed in Section 4 in terms of jet precession due to 
   the Bardeen-Petterson effect. General conclusions are summarized in Section 5.

\section{The Bardeen-Petterson effect}

\subsection{The Bardeen-Petterson radius}

   Frame dragging produced by a Kerr black hole causes precession of a 
   particle if its orbital plane is inclined in relation to the equatorial 
   plane of the black hole. This effect, known as Lense-Thirring precession, 
   has been mainly studied analytically and numerically in the context of 
   quasi-periodic X-ray brightness oscillations in X-ray binaries 
   \citep{stvi98,mala98,arna99}.
   The precession angular velocity $\Omega_{\mathrm{LT}}$ 
   due to the Lense-Thirring effect is given by (e.g., \citealt{wilk72}):

   \begin{eqnarray}
      \Omega_\mathrm{LT}(r) = \frac{2G}{c^2}\frac{J_\mathrm{BH}}{r^3},
   \end{eqnarray}
   \\where $G$ is the gravitational constant, $c$ is the speed of light, $r$ 
   is the distance from the black hole and $J_\mathrm{BH}$ is the angular 
   momentum of a Kerr black hole with mass $M_\mathrm{BH}$, which is 
   defined as: 

   \begin{eqnarray}
      J_\mathrm{BH} = a_\ast \frac{GM_\mathrm{BH}^2}{c},
   \end{eqnarray}
   \\where $a_\ast$ ($|a_\ast| \leq 1$) is a dimensionless parameter 
   corresponding to the ratio between the actual angular momentum of the 
   black hole and its maximum possible value. 

   The alignment between the angular momenta of the Kerr black hole and the 
   accretion disc is forced by a combination between the Lense-Thirring 
   effect and the internal viscosity of the accretion disc. This is known as the Bardeen-Petterson 
   effect \citep{bape75} and it tends to affect the innermost part of the disc due to 
   the short range of the Lense-Thirring effect, while the disc's outer part tends to remain 
   in its original configuration. The transition radius between these two regimes 
   is known as the Bardeen-Petterson radius $R_\mathrm{BP}$ and its exact location 
   depends mainly on the physical properties of the accretion disc 
   (\citealt{bape75,kupr85,scfe96,ivil97,arpr98,nepa00,lub02,fran05}). Basically, 
   the Bardeen-Petterson radius is determined by comparing the time-scale related 
   to the Lense-Thirring effect to that of warp transmission through the disc, giving:

   \begin{eqnarray}
      R_\mathrm{BP}^\mathrm{diff}=\sqrt{\frac{\nu_2(R_\mathrm{BP}^\mathrm{diff})}
        {\Omega_\mathrm{LT}(R_\mathrm{BP}^\mathrm{diff})}},
   \end{eqnarray} 
   \\or

   \begin{eqnarray}  
      R_\mathrm{BP}^\mathrm{w} = \frac{c_\mathrm{s}(R_\mathrm{BP}^\mathrm{w})}
         {\Omega_\mathrm{LT}(R_\mathrm{BP}^\mathrm{w})},
   \end{eqnarray}
   \\depending on whether the warp propagation occurs diffusively ($R_\mathrm{BP}^\mathrm{diff}$) 
   or through waves ($R_\mathrm{BP}^\mathrm{w}$). In equations (3) and (4), 
   $\nu_2$ represents the viscosity along the normal to the accretion disc, 
   and $c_\mathrm{s}$ is the sound speed in the disc.
 
   \citet{pali95} showed that the transition from the diffusive to wave-like 
   regime occurs at a radius $R_\mathrm{T}\sim H_\mathrm{d}/\alpha$, where 
   $H_\mathrm{d}$ is the scale-height of the disc and $\alpha$ is the 
   dimensionless viscosity parameter introduced by \citet{shsu73}.

\subsection{Alignment time-scale}

   The time-scale for alignment between the angular momenta of the black hole 
   and the accretion disc was first estimated by \citet{rees78}, assuming 
   that each mass element accreted by the black hole carries an orbital angular 
   momentum corresponding to that at the Bardeen-Petterson radius. 
   \citet{scfe96} obtained an analytic solution to the equations that control 
   the warp evolution in the case of a disc with constant surface density and 
   calculated the alignment time-scale. \citet{naar99} generalized the results 
   found by \cite{scfe96} to a power-law viscosity. These studies suggest that 
   the alignment time-scale can be estimated by:

   \begin{eqnarray}
      T_\mathrm{align} =
       J_\mathrm{BH}\left(\frac{dJ_\mathrm{BH}}{dt}\right)^{-1}\sin\varphi,
   \end{eqnarray}
   \\where $\varphi$ is the angle between the black hole spin axis and the 
   direction perpendicular to the outer part of the warped disc. The time 
   derivative of $J_\mathrm{BH}$ has the form:

   \begin{eqnarray}
      \frac{dJ_\mathrm{BH}}{dt} =
       -2\pi\sin\varphi\int_{R_\mathrm{BP}}^{R_\mathrm{out}}\Omega_\mathrm{LT}(r)L_\mathrm{d}(r)rdr,
   \end{eqnarray}
   \\where $L_\mathrm{d}$ is the angular momentum density of the accretion disc, 
   given by:

   \begin{eqnarray}
      L_\mathrm{d}(r) = \Sigma(r)\Omega_\mathrm{d}(r)r^2, 
   \end{eqnarray}
   \\where $\Omega_\mathrm{d}$ and $\Sigma$ are respectively the angular 
   velocity and the mass surface density of the accretion disc. The last 
   quantity is obtained by integrating over the scale-height of the disc:

   \begin{eqnarray}
      \Sigma(r) = \int_{-H_\mathrm{d}/2}^{H_\mathrm{d}/2}\rho(r,z)dz,
   \end{eqnarray}
   \\where $\rho$ is the mass density of the disc and $z$ is the perpendicular 
   distance from the midplane of the disc. 

   Note that the alignment torque depends on an integral that should 
   be performed from the Bardeen-Petterson radius to the outer radius 
   of the accretion disc. Although the last quantity is not known accurately 
   in many cases (even in the case of NGC\,4258), its precise value is 
   not important for our calculations since the alignment torque decreases 
   rapidly with distance for the realistic disc models considered in this 
   work.

\section{Accretion disc of NGC\,4258}

   The examination of the viability of the Bardeen-Petterson 
   effect requires some knowledge about the physical characteristics 
   of the accretion disc (such as its viscosity, surface density and 
   scale-height), as well as the mass and spin of the black hole. 
   In this section, we will present our parametric accretion disc 
   model for NGC\,4258, using the available observational data to 
   constrain the values of the parameters.

\subsection{Parametric model for the accretion disc}

   We will consider an accretion disc with total 
   mass $M_\mathrm{d}$. In the case of an axisymmetric disc, 
   its total mass can be calculated through:

   \begin{eqnarray}
      M_\mathrm{d}(r\leq R_\mathrm{out}) = 2\pi\int_{R_\mathrm{ms}}^{R_\mathrm{out}}\Sigma(r)rdr.
   \end{eqnarray}

   Here $R_\mathrm{out}$ is the outer radius of the 
   accretion disc, while $R_\mathrm{ms}=A_\mathrm{ms}(a_\ast)R_\mathrm{g}$ 
   is the radius of the marginally stable orbit (assumed to 
   be the inner radius of the disc), $R_\mathrm{g}=GM_\mathrm{BH}/c^2$ 
   is the gravitational radius, and 
   $A_\mathrm{ms}(a_\ast) = 3+A_2\mp\sqrt{(3-A_1)(3+A_1+2A_2)}$, with 
   $A_1 = 1+(1-a_\ast^2)^{1/3}\left[(1+a_\ast)^{1/3}+(1-a_\ast)^{1/3}\right]$ 
   and $A_2 = \sqrt{3a_\ast^2+A_1^2}$ (e.g., \citealt{bar72}). In this case, 
   the minus and plus signs correspond to prograde and  retrograde motion, 
   respectively. 

   The scale-height of the disc $H_\mathrm{d}$ can be estimated following \citet{saco81} (see also 
   \citealt{cap06}):

   \begin{eqnarray}
      H_\mathrm{d}(r) = 
       2\frac{H_\mathrm{nsg}(r)H_\mathrm{sg}(r)}{\sqrt{H_\mathrm{nsg}^2(r)+H_\mathrm{sg}^2(r)}},
   \end{eqnarray}
   \\where 

   \begin{eqnarray}
      H_\mathrm{nsg}=c_\mathrm{s}/\Omega_\mathrm{d}, 
   \end{eqnarray}
   \\and

   \begin{eqnarray}
      H_\mathrm{sg}=c_\mathrm{s}^2/(\pi G\Sigma_\mathrm{d}).
   \end{eqnarray}

   This formulation takes into account possible effects related to self-gravity. 
   In the case of NGC\,4258, maser observations have shown that this effect is 
   negligible (e.g., \citealt{her05}), so that we will assume hereafter 
   $H_\mathrm{d}=H_\mathrm{nsg}$.

   The sound speed $c_\mathrm{s}$ is given by (e.g., \citealt{abr88}):

   \begin{eqnarray}
      c_\mathrm{s}(r) = \sqrt{-\Gamma\frac{d\ln\Omega_\mathrm{d}(r)}{d\ln r}
       \frac{\nu_1(r)\Omega_\mathrm{d}(r)}{\alpha}},
   \end{eqnarray}
   \\where $\Gamma$ is the politropic index of the gas, which we have taken 
   to be equal to 5/3, and $\nu_1$ is the viscosity along the disc:

   \begin{eqnarray}
      \nu_1(r) = -\frac{\dot{M}}{2\pi\Sigma_\mathrm{d}(r)}\left[
        \frac{d\ln\Omega_\mathrm{d}(r)}{d\ln r}\right]^{-1}\left[1-\left(\frac{R_\mathrm{ms}}{r}\right)^2
        \frac{\Omega_\mathrm{d}(R_\mathrm{ms})}{\Omega_\mathrm{d}(r)}\right],
   \end{eqnarray}
   \\where $\dot{M}$ is the accretion rate onto the black hole, 
   assumed to be constant along the disc.

   The viscosity acting in the vertical direction, $\nu_2$, is written 
   in terms of $\nu_1$ as $\nu_2\sim f(\alpha)\nu_1$, where 
   $f(\alpha)=2(1+7\alpha^2)/[\alpha^2(4+\alpha^2)]$ \citep{ogil99}. 
   If the warp propagates diffusively along the disc in the linear 
   regime and $\alpha\ll 1$, we have $f(\alpha)\approx 1/2\alpha^2$ 
   \citep{papr83}. Since $f(\alpha)$ is substantially greater than unity 
   for the usual values of $\alpha$, warping modes very effectively propagate 
   inward the disc, where they are dissipated later. Note that in the 
   strong non-linear regime (warps with very large amplitudes), additional 
   dissipation caused by fluid instabilities might reduce $f(\alpha)$, so 
   that $\nu_2/\nu_1\approx 1$ \citep{gam00}. Both possibilities were studied 
   by \citet{lopr06} in their numerical simulations of the Bardeen-Petterson 
   effect. In this work, we are assuming that the linear regime is applicable 
   for the accretion disc of NGC\,4258.

\subsection{Constraining accretion disc model parameters for NGC\,4258 from maser observations}

\subsubsection{Mass of the black hole and accretion disc rotation law}

   The black hole mass in NGC\,4258 has been determined from interferometric maser 
   observations \citep{miy95,mor99,her05}; we will adopt $M_\mathrm{BH}=3.78\times 10^7 M_{\sun}$, 
   as inferred by \citet{her05}.

   The rotation curve of the maser spots is compatible with Keplerian motions (e.g., 
   \citealt{miy95,her05}). Therefore, we will assume a (relativistic) Keplerian angular 
   velocity $\Omega_\mathrm{K}$ for the accretion disc of the form:

   \begin{eqnarray}
      \Omega_\mathrm{K}(r) = \frac{c^3}{GM_\mathrm{BH}}\left[\left(\frac{r}{R_\mathrm{g}}\right)^{3/2}+a_\ast \right]^{-1}.
   \end{eqnarray}

%-----------------------------------------------------------FIGURE 01 
   \begin{figure}
      {\includegraphics{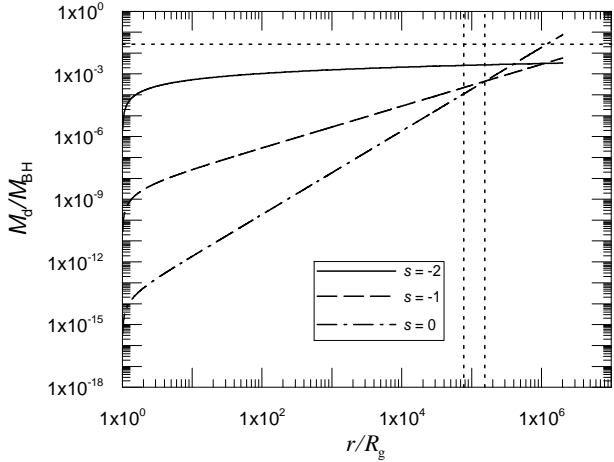}}
      \caption{Cumulative mass of the accretion disc of NGC\,4258, normalized by the black hole 
      mass, as a function of radius for the upper limit of $\Sigma_0$. The continuous line 
      refers to a mass surface density of the disc decreasing with the square of the distance ($s=-2$), 
      while the dashed and dashed-dotted lines correspond respectively to $s=-1$ and 0. The 
      vertical dashed lines mark the inner and outer radii of the maser disc. The horizontal 
      dashed line shows the observed upper limit for $M_\mathrm{d}/M_\mathrm{BH}$ \citep{mor99}. 
      }
      \label{disc mass}
   \end{figure}

%______________________________________________________________
 
\subsubsection{Surface density of the accretion disc}

   We have adopted a power-law radial distribution for the accretion disc 
   surface density (e.g., \citealt{pate95,larw97,nepa00}),

   \begin{eqnarray}
      \Sigma(r)=\Sigma_0\left (\frac{r}{R_\mathrm{g}}\right )^s,
   \end{eqnarray}
   \\where $\Sigma_0$ and $s$ are constants to be determined from observations.

   From the relation between mass and surface densities of the accretion disc 
   (equation 8), $\Sigma_0$ can be expressed through:

   \begin{eqnarray}
      \Sigma_0=\rho R_\mathrm{g}\left (\frac{H_\mathrm{d}}{r}\right )\left (\frac{r}{R_\mathrm{g}}\right )^{1-s},
   \end{eqnarray}
   \\where $\rho=\mu m_\mathrm{H}n$, $\mu$ is the mean molecular weight, and 
   $m_\mathrm{H}$ is the mass of the hydrogen atom.

   The maser amplification is favoured when the particle density $n$ 
   ranges from $10^8$ to $10^{10}$ cm$^{-3}$ \citep{elit92}. Thus, for 
   NGC\,4258, this condition must be fulfilled. A statistical analysis 
   of the vertical structure in the systemic masers of NGC\,4258 
   has suggested that $H_\mathrm{d}/r\la0.002$ \citep{her05}. Taking 
   into account this observational constraint, equation (17) provides an 
   upper limit on $\Sigma_0$, such that 
   $\Sigma_0^\mathrm{max}(s)\la1.85\times 10^{-4}(r=0.14\,\mathrm{pc}/R_\mathrm{g})^{1-s}$ g cm$^{-2}$, 
   calculated at the inner radius of the maser disc for a particle density of $10^{10}$ 
   cm$^{-3}$. 

   On the other hand, using equation (11) for the scale-height of the 
   disc, we obtain:

   \begin{eqnarray}
      \Sigma_0 = \left(\frac{\Gamma}{2\pi R_\mathrm{g}}\right)\left(\frac{\dot{M}}{\alpha}\right)\left(\frac{H_\mathrm{d}}{r}\right)^{-2}\left(
      \frac{r}{R_\mathrm{g}}\right)^{-(s+1/2)}.
   \end{eqnarray}
   
   Using again the observational constraint $H_\mathrm{d}/r\la0.002$ 
   at $r=0.14\,\mathrm{pc}$ in equation (18), we obtain a lower limit 
   for $\Sigma_0$, denoted by $\Sigma_0^\mathrm{min}$.

%-----------------------------------------------------------TABLE 01 

\begin{table}
 \centering
 \begin{minipage}{100mm}
  \caption{Accretion disc parameters for NGC\,4258.}
  \begin{tabular}{@{}ccccc@{}}
  \hline
  s & $\Sigma_0$ \footnote{Upper limit.} & 
$\epsilon\dot{M}$ \footnote{Highest value among all black hole spins ($\epsilon\dot{M}\geq 1.77\times 10^{-6}$ M$_\odot$ yr$^{-1}$).} & 
$\alpha$ \footnote{Lowest value among all black hole spins ($\alpha\leq 0.2$).} & 
$L_\mathrm{bol}/L_\mathrm{Edd}$ \footnote{Highest value among all black hole spin values ($L_\mathrm{bol}/L_\mathrm{Edd}\geq 2.10\times 10^{-5}$).}
\\
 & (g cm$^{-2}$) & (M$_\odot$ yr$^{-1}$) & & \\
 \hline
-2 & $8.60\times 10^{10}$ & $1.34\times 10^{-5}$ &  0.026 & $1.60\times 10^{-4}$\\
-1 & $1.11\times 10^{6}$  & $1.34\times 10^{-5}$ &  0.026 & $1.60\times 10^{-4}$\\
 0 & $1.43\times 10^{1}$  & $1.34\times 10^{-5}$ &  0.026 & $1.60\times 10^{-4}$\\
\hline
\end{tabular}
\end{minipage}
\end{table}

%______________________________________________________________

%-----------------------------------------------------------FIGURE 02 
   \begin{figure*}
      {\includegraphics{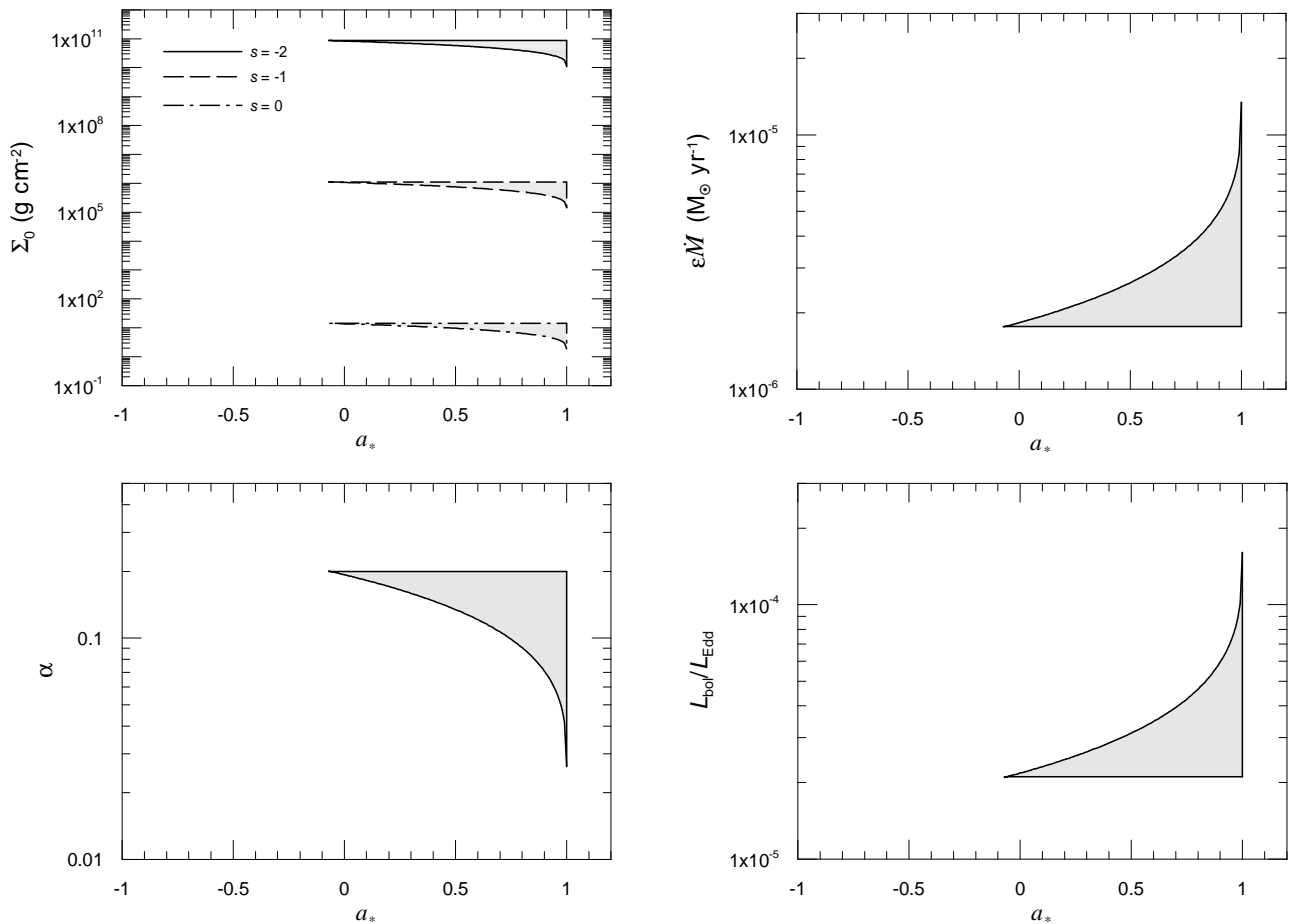}}
      \caption{Accretion disc parameters of NGC\,4258. {\it Upper panels}: The surface density at 
      $r=R_\mathrm{g}$ for three different power-law dependencies of the disc surface density 
      ($s=-2, -1$ and 0) and the accretion rate multiplied by the accretion efficiency 
      as a function of the black hole spin. The gray areas show the range of parameters that obey the 
      observational constraints. {\it Lower panels}: The alpha parameter and the bolometric luminosity in units of the 
      Eddington limit as a function of the black hole spin (same nomenclature as in the previous panels is 
      used).
      }
      \label{Disc parameters}
   \end{figure*}

%______________________________________________________________

   Note that this equation was obtained under the assumption that 
   viscosity is responsible for disc heating. However, the warped maser 
   region is also heated by the X-ray continuum from the central 
   source \citep{nema95}, which increases the magnitude of the local sound 
   speed relative to that calculated from equation (13). Consequently, 
   it could change the value of $\Sigma_0^\mathrm{min}$ calculated via 
   equation (18). In this paper we will neglect the X-ray heating since 
   its inclusion does not change substantially either our estimates of the 
   Bardeen-Petterson radius or the associated alignment timescale. 
   The use of an isothermal model as in \citet{nema95} would change the 
   Bardeen-Petterson radius by a factor in a range of about 0.2 - 2, while for 
   the alignment timescale the correction would not be larger than a 
   factor of two.

   From the upper limit of $10^6 M_{\sun}$ for the disc mass inferred 
   from the maser observations \citep{mor99,her05}, we can also establish 
   a lower limit on $s$. Substituting the upper limit on $\Sigma_0$ 
   determined above into equation (9), we found that only for $s\ga-2.3$ 
   the upper limit on the the disc mass is satisfied. As an example, 
   we present in Figure 1 the cumulative mass of the accretion disc 
   as a function of radius, considering $s=-2$, $-1$ and 0. Note that 
   this quantity resides below the observational upper limit for 
   the disc mass of NGC\,4258 in all three cases.

\subsubsection{Accretion rate, viscosity and bolometric luminosity}

  The bolometric luminosity of a source is related to the mass accretion rate through: 

   \begin{eqnarray}
      L_\mathrm{bol} = \epsilon\dot{M}c^2,
   \end{eqnarray} 
   \\where 

   \begin{eqnarray}
      \epsilon = 1-\frac{1-2A_\mathrm{ms}^{-1}+a_\ast A_\mathrm{ms}^{-3/2}}{\sqrt{1-3A_\mathrm{ms}^{-1}+2a_\ast A_\mathrm{ms}^{-3/2}}} 
   \end{eqnarray}
   \\is the accretion efficiency onto a Kerr black hole (e.g., \citealt{shte83}).

   The bolometric luminosity of NGC\,4258 is roughly known, ranging approximately 
   from $10^{41}$ to $10^{43}$ erg s$^{-1}$, which translates to 

   \begin{eqnarray}
      \frac{1.765\times 10^{-6}}{\epsilon}\la\dot{M} \,\,(\mathrm{M}_\odot\,\mathrm{yr}^{-1})\la \frac{1.765\times 10^{-4}}{\epsilon}. 
   \end{eqnarray}

   However, there is a more restrictive upper limit for 
   $\dot{M}$ since the condition $\Sigma_0^\mathrm{min}\la\Sigma_0^\mathrm{max}$ must 
   be always satisfied. This relation leads to 

   \begin{eqnarray}
      \dot{M}\la 1.60\times 10^{-4}\alpha \,\,\mathrm{M}_\odot\,\mathrm{yr}^{-1},
   \end{eqnarray}
   \\which is compatible with the recent estimates obtained by \citet{her05} 
   and \citet{mod05}, as well as with the accretion rate inferred previously 
   by \citet{nema95}. It is important to emphasize that this 
   limit does not depend on the value of the power-law index $s$.

   On the other hand, the upper limit of $\dot{M}$ obtained from equation (22) must not be lower than 
   its minimum value calculated from equation (21), implying that $\epsilon\alpha\ga 0.011$, 
   which translates to a lower limit for the $\alpha$ parameter ($\alpha\ga 0.026$ for $\epsilon\la 0.42$). 
   Numerical magneto-hydrodynamic simulations, as well as observations of many types of 
   accretion disc systems have indicated that $0.01\la\alpha\la 0.2$ (\citealt{sicz89,bra95,haw96,sto96,smak99,men00,hakr01,win03}; 
   and see \citealt{kin07} for a discussion).    

   Regarding the black hole spin, the conditions $\alpha\la 0.2$ and $\epsilon\alpha\ga 0.011$ 
   are fulfilled only if $a_\ast\ga-0.07$ in equation (20), which might be suggesting that the 
   supermassive black hole in NGC\,4258 is rotating progradely in relation to its 
   accretion disc unless it is spinning very slowly. 

   In Figure 2 we show the parameters $\Sigma_0$, $\epsilon\dot{M}$, $\alpha$ and $L_\mathrm{bol}/L_\mathrm{Edd}$, 
   where $L_\mathrm{Edd}$ is the Eddington luminosity, as a function of the 
   black hole spin for $s=$-2, -1 and 0. 

   In Table 1 we summarize the physical parameters of NGC\,4258 that will be used to 
   analyze the feasibility of the Bardeen-Petterson effect in this particular source.

\section{Bardeen-Petterson effect in NGC\,4258}

\subsection{Bardeen-Petterson radius}

   Considering the extreme values of $\Sigma_0$, $\alpha$ 
   and $\dot{M}$ for $s=-2, -1$ and 0 (see Figure 2), we found in all cases that 
   $R_\mathrm{T}\gg10^7 R_\mathrm{ms}$, much greater than the outer radius of the 
   maser disc. Therefore, if our disc parameterization is correct, the accretion 
   disc of NGC\,4258 responds diffusively to the Bardeen-Petterson effect, which 
   implies that we must use equation (3) to obtain $R_\mathrm{BP}$. As we mentioned 
   in section 2, we are assuming that $\nu_2\sim f(\alpha)\nu_1$; plugging it into 
   equation (3), we were therefore able to determine the Bardeen-Petterson radius for each 
   set of disc parameters.

   In Figure 3 we show the Bardeen-Petterson radius as a function of the black hole 
   spin for NGC\,4258 for three different power-law surface mass density distribution 
   ($s=-2, -1$ and 0). We have also considered four distinct cases related to different 
   combinations among the parameters $\Sigma_0$, $\alpha$ and $\dot{M}$ in order to 
   produce the extreme values for $R_\mathrm{BP}$. 

   We can see that the values of the Bardeen-Petterson radius are comparable to or 
   smaller than 0.14 pc, the inner radius of the maser disc inferred by \citet{her05}. 
   At first glance, this result may be seen as arguing against the Bardeen-Petterson 
   effect since the warp has been inferred observationally in the outer disc. However, numerical 
   simulations of Bardeen-Petterson discs, for several distinct physical conditions, 
   usually indicate a smooth transition between the inner (flat) Bardeen-Petterson 
   disc and the outer (misaligned) disc, exhibiting a gradient in the inclination angle 
   \citep{nepa00,fran05,lopr06}. Consequently, warping in the outer disc could 
   be associated with the gradient in the disc tilt angle, produced by the transition 
   between a Bardeen-Petterson disc, and a misaligned outer disc that includes the 
   maser region.

   Indeed, the best-fitting model for the maser kinematics is a warped disc with a 
   radial inclination gradient \citep{her05}. This model leads to a difference 
   in inclination of about eight degrees between the inner and outer maser 
   radius. Considering the full range of values for the Bardeen-Petterson radius 
   displayed in Figure 3, and extrapolating the inclination gradient to the inner disc, 
   we have found a variation in the tilt angle at the Bardeen-Petterson radius and innermost 
   part of the maser disc that ranges from four to eight degrees approximately. 
   
   Besides the warp in the tilt angle, there is also strong evidence that 
   the maser disc of NGC\,4258 is twisted azimuthally, in the sense that the position 
   angle in the disc varies with the radial distance from the black hole (e.g., 
   \citealt{her05} and references therein). Indeed, \citet{her05} found that a 
   variation of about 10$\degr$ in the warp position angle is sufficient to accommodate 
   the observed maser kinematics. Assuming that the warping in the position angle 
   is described by 
   the quadratic function given by these authors, we find a variation of $\sim 17\degr$ 
   in this quantity from the inner maser radius to the extrapolated value at $r\sim 0$ mas. 
   Interestingly, a disc twisting of about 10$\degr$ is obtained if we consider the 
   variation of the position angle from the core to the highest value of the Bardeen-Petterson 
   radius calculated in this work. \citet{cec00} and \citet{wil01} proposed that jet 
   precession by an angle of about 10$\degr$ can also reproduce the physical characteristics 
   of the anomalous arms. The similarity between the required precession and warp position 
   angle seems to suggest that they could be arising from the same physical mechanism. 
   As we shall discuss in Section 4.3, the Bardeen-Petterson effect can provide the 
   required jet precession in NGC\,4258. We therefore propose the same mechanism to be 
   responsible for the azimuthal twist of the disc.

   The allowed range for $R_\mathrm{BP}$ decreases 
   for smaller values of the black hole spin, maintaining this trend even for retrograde cases 
   (as a consequence of the more restrictive limits on the disc parameters 
   in those cases).

%-----------------------------------------------------------FIGURE 03 
   \begin{figure}
      {\includegraphics{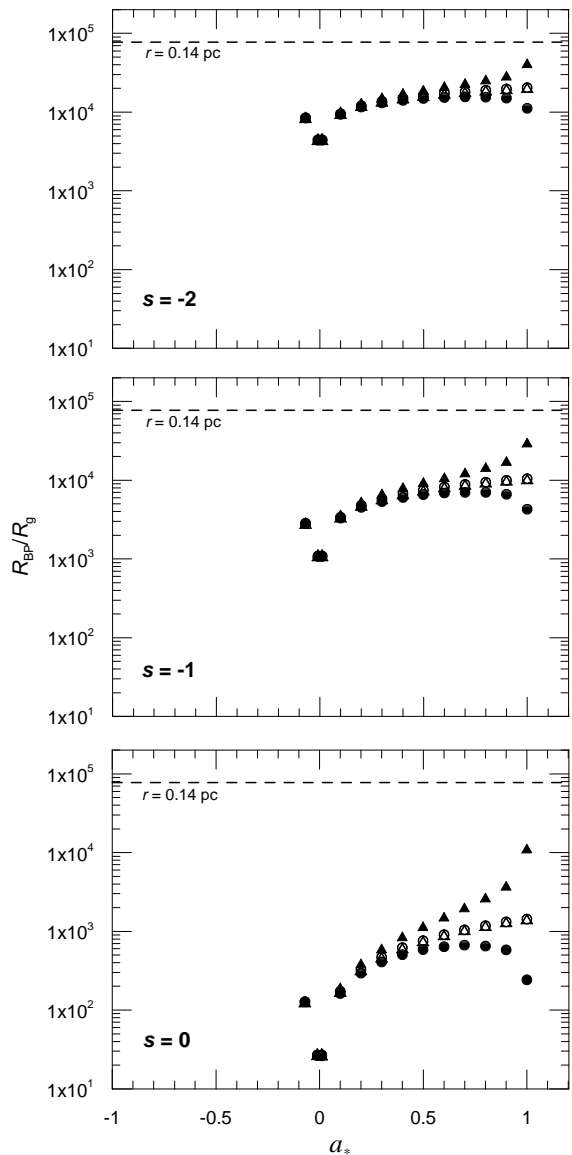}}
      \caption{The Bardeen-Petterson radius as a function of the black hole spin. From the upper 
      to the lower panel, $s$ varies from -2 to 0. Open circles correspond to the case of 
      minimum value for $\Sigma_0$ and $\dot{M}$, and $\alpha=0.2$ (the maximum value considered 
      in this work). Full circles represent the maximum value for $\Sigma_0$ and the 
      minimum for $\dot{M}$ and $\alpha$. Full triangles refer to the case of minimum value 
      for $\dot{M}$ and maximum for $\Sigma_0$ and $\alpha$, while open triangles represent 
      the maximum for the three disc parameters. The dashed lines mark the position of the 
      inner radius of the maser disc ($r=0.14$ pc).
      }
      \label{Bardeen-Petterson radius}
   \end{figure}

%______________________________________________________________

   Comparing the solutions represented by the full triangles and the full circles in 
   Figure 3, we can infer the influence of $\alpha$ on $R_\mathrm{BP}$. Fixing $s$ 
   and $a_\ast$, the Bardeen-Petterson radius is larger when $\alpha$ is at maximum, 
   increasing the coupling between consecutive disc annuli and consequently the efficiency 
   of the Bardeen-Petterson effect. In fact, we can see from equation (3) that 
   $R_\mathrm{BP} \propto f(\alpha)^{1/(s-1)}$, for $R_\mathrm{BP}\gg R_\mathrm{g}$.

   The influence of $\Sigma_0$ on $R_\mathrm{BP}$ can be examined from a comparison
   between the open circles and full triangles: the Bardeen-Petterson radius 
   increases with $\Sigma_0$, which reflects the fact that 
   $R_\mathrm{BP} \propto \Sigma_0^{-1/(s-1)}$. A comparison between the solutions 
   represented by the full and open triangles shows that a decrease in the accretion 
   rate results in an expansion of the Bardeen-Petterson radius, since 
   $R_\mathrm{BP} \propto \dot{M}^{1/(s-1)}$.

\subsection{Alignment time-scale due to the Bardeen-Petterson effect}

   In Figure 4 we present the alignment time-scale as a function of the 
   black hole spin, calculated from equation (5). For a fixed set of disc 
   parameters, we can see a clear anti-correlation between $T_\mathrm{align}$ 
   and $s$ (in the sense that as $s$ increases, the alignment time-scale 
   becomes shorter), which is a consequence of the inertia of the disc (the 
   disc mass inside the Bardeen-Petterson radius decreases as the value of 
   $s$ increases, as shown in Figure 1). The influence of $\alpha$, 
   $\Sigma_0$ and $\dot{M}$ on the alignment timescale can be analysed 
   (in a similar fashion to our discussion of the Bardeen-Petterson radius) 
   by comparing the different symbols displayed in Figure 4 
   ($T_\mathrm{align} \propto \Sigma_0^{-1}R_\mathrm{BP}^{(1/2-s)}$, 
   for $R_\mathrm{BP}\gg R_\mathrm{g}$).

%-----------------------------------------------------------FIGURE 04 
   \begin{figure}
      {\includegraphics{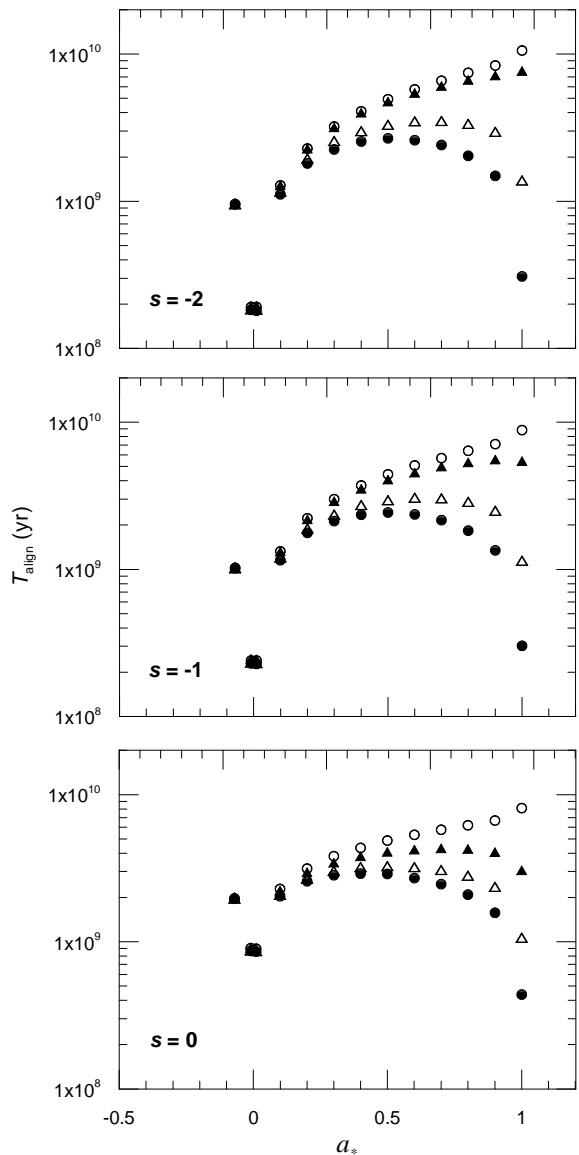}}
      \caption{The alignment time-scale as a function of the black hole spin for NGC\,4258. From the upper 
      to the lower panel, $s$ varies from -2 to 0. As in Figure 3, open circles correspond to the 
      case of minimum value for $\Sigma_0$ and $\dot{M}$, and $\alpha=0.2$, the maximum value considered 
      in this work. Full circles represent the maximum value for $\Sigma_0$ and the 
      minimum for $\dot{M}$ and $\alpha$. Full triangles refer to the case of minimum value 
      for $\dot{M}$ and maximum for $\Sigma_0$ and $\alpha$, while open triangles represent 
      the maximum for the three disc parameters.
      }
      \label{Alignment Time-scale}
   \end{figure}

%______________________________________________________________

   Our results show that the Bardeen-Petterson effect acting upon the 
   accretion disc of NGC\,4258 leads to alignment time-scales which are 
   typically in the range 
   from 0.4 to 10 Gyr.

\subsection{Anomalous arms and the radio jet of NGC\,4258}

   The anomalous arms, observed from radio to X-ray wavelengths 
   (e.g., \citealt{bur63,kru72,cec00,wil01}), are strongly believed 
   to lie in the galactic disc of NGC\,4258 (e.g., \citealt{kru74,alhu82,wil01}), 
   even though their nature has been related to the jet propagation 
   through the disc in some cases \citep{for86,mar89,cec92}. The 
   anomalous arms are straight at the inner parts of the disc and 
   show, after about 2 kpc, a gradual curvature in opposite directions 
   in relation to the nuclear region, forming a S-shape structure.

   \citet{wil01} performed a detailed study of NGC\,4258, using 
   high-resolution X-ray observations, correlating these data with 
   radio and optical images obtained in previous works. They 
   proposed that the anomalous arms are shocked regions produced 
   by the interaction between the disc gas and the mass motions 
   induced by the propagation of the radio jet/counterjet. In 
   addition, they suggested two different scenarios for the 
   curvature of the anomalous arms: jet precession, also proposed 
   by \citet{cec00}, or buoyancy of the less dense and less 
   tightly bound gas in the outer galactic disc. 

   Here we shall assume that the curvature of the outer 
   anomalous arms of NGC\,4258 is driven by jet precession. From 
   this assumption, we shall analyze the possibility that this 
   precession is being produced by the Bardeen-Petterson effect. 

   The Bardeen-Petterson effect induces not only a disc alignment 
   but also a precession around the rotation axis of the black hole 
   (e.g., \citealt{scfe96,cap04,cap06}). Indeed, \citet{scfe96} showed that 
   the alignment and precession time-scales are identical, with the 
   precession period varying exponentially with time:

   \begin{eqnarray}
      P_\mathrm{prec}(\Delta t)=T_\mathrm{align}e^{-\Delta t/T_\mathrm{align}},
   \end{eqnarray}
   \\where $P_\mathrm{prec}$ is the precession period after an elapsed 
   time interval $\Delta t$.

   The precession angle $\Delta\vartheta$ as a function of $\Delta t$ 
   and $T_\mathrm{align}$ can be written as:

   \begin{eqnarray}
      \Delta\vartheta(\Delta t)=2\pi\frac{\Delta t}{T_\mathrm{align}}e^{\Delta t/T_\mathrm{align}}.
   \end{eqnarray}

   In order to be able to reproduce the physical characteristics of the anomalous 
   arms, as well as the radio jet, it is necessary for the jet direction 
   to change by $\sim 10\degr$ \citep{cec00,wil01}, which 
   implies that $\Delta\vartheta\approx 10\degr$. Therefore, we can use 
   equation (24) to estimate the time interval required to obtain a precession 
   of the jet by ten degrees $\Delta t_\mathrm{10\,deg}$, 
   for each solution derived in the previous sections. The results are 
   shown in Figure 5.

%-----------------------------------------------------------FIGURE 05 
   \begin{figure}
      {\includegraphics{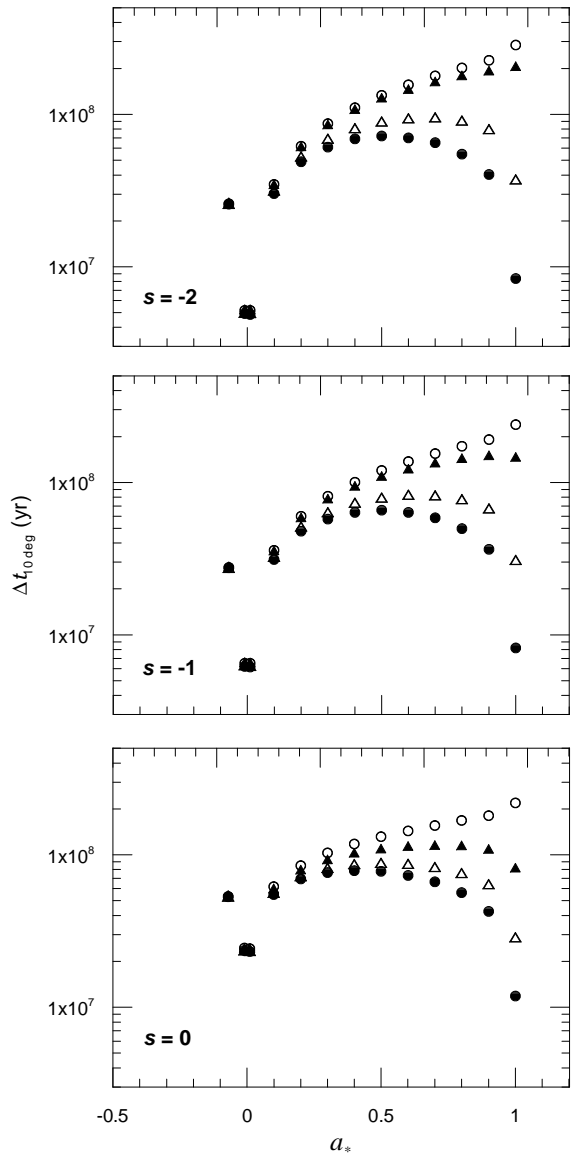}}
      \caption{Time interval for the jet/accretion disc of NGC\,4258 to precess by an angle of 10 
      degrees due to the Bardeen-Petterson effect as a function of the black hole spin. Symbols 
      have the same meaning of those plotted in Figure 4.
      }
      \label{Alignment Time-scale}
   \end{figure}

%______________________________________________________________

   We note that a time interval between 5 and 300 Myr is necessary 
   for the jet to change its direction by ten degrees. Such time-scales 
   seem to be feasible since they are similar to or shorter than the viscous 
   time-scale at 0.14 pc (between 0.06 and 3 Gyr considering all combinations 
   of the disc parameters used here). Therefore, the Bardeen-Petterson 
   effect is an excellent candidate for driving the precession in NGC\,4258.

\section{Conclusions}

   In this work, we have analysed the viability of the Bardeen-Petterson 
   effect as the mechanism responsible for the warping and precession of 
   the disc of NGC\,4258. Evidence for a warped accretion disc in this 
   source comes from the kinematics of the water maser inferred from 
   high-resolution observations \citep{miy95,her05}. Precession of 
   the radio jet/counterjet has been pointed as the potential reason 
   for the curvature of the outer anomalous arms in the galactic disc of 
   NGC\,4258 \citep{cec00,wil01}.

   Assuming a power-law surface density distribution for the accretion 
   disc material, we used constraints inferred from observations and 
   determined the acceptable ranges for the values of the accretion 
   rate and the surface density at the inner radius of the disc. Our 
   results also show a dependency of the black hole spin parameter on 
   the power-law index of the surface density, such that $a_\ast$ cannot 
   be smaller than about -0.07 if the observational constraints are to 
   be obeyed.

   We have calculated the Bardeen-Petterson radius for a variety of disc 
   parameters and, in all cases, it is similar to or located inside of the inner 
   radius of the maser disc. This argues for the existence of 
   a Bardeen-Petterson disc in NGC\,4258. Consequently, the 
   observed warping may be interpreted as resulting from a gradient in 
   the disc inclination angle due to the transition from an inner 
   (flat) disc and a misaligned outer disc. All solutions lead to 
   a time-scale for the alignment between the accretion disc within 
   the Bardeen-Petterson radius and the black hole's equator shorter 
   than about 10 Gyr. The alignment time-scales obtained were generally 
   longer than about 0.4 Gyr.

   Assuming that the time evolution of the precession period follows 
   the results obtained by \citet{scfe96}, we calculated the time interval 
   necessary for the accretion disc, and consequently for the jet/counterjet, 
   to precess by an angle of ten degrees. This value is necessary 
   to reproduce the curvature of the outer anomalous arms in the kiloparsec 
   galactic disc of NGC\,4258 if jet precession is responsible for it. We 
   found that an interval between 5 and 300 Myr can make the jet to change 
   its orientation by the required angle. This time-scale is shorter 
   than the radial viscous time-scale calculated at 0.14 pc (between 0.06 
   and 3 Gyr, depending on the disc parameters), the inner radius of the 
   maser disc. 
   Given the similarity between the value of the jet precession 
   angle and that of the twist of the azimuthal position angle inside the 
   Bardeen-Petterson region of the disc, we suggest the possibility that 
   both effects are associated with the same physical origin (frame-dragging 
   by a Kerr black hole).

   We conclude that the Bardeen-Petterson effect may explain the 
   observations of NGC\,4258 on both the sub-parsec and kiloparsec 
   scales.

\section*{Acknowledgments}

      This work was supported by the Brazilian Agencies FAPESP and CNPq. 
A.C. acknowledges the hospitality of the Space Telescope Science Institute, 
where this work was partially carried out. We acknowledge very helpful 
remarks from an anonymous referee.

\bsp

\label{lastpage}


\begin{thebibliography}{99}

\bibitem[Abramowicz et al.(1988)]{abr88} Abramowicz, M. A., Czerny, B., Lasota, J. P., Szuszkiewicz, E. 1988, 
       ApJ, 332, 646

\bibitem[Armitage \& Natarajan(1999)]{arna99} Armitage, P. J., Natarajan, P. 1999, 
       ApJ, 525, 909

\bibitem[Bardeen, Press \& Teukolsky(1972)]{bar72} Bardeen, J. M., Press, W. H., Teukolsky, S. A. 1972, 
      ApJ, 178, 347

\bibitem[Bardeen \& Petterson(1975)]{bape75} Bardeen, J. M. \& Petterson, J. A. 1975, 
      ApJ, 195, L65

\bibitem[Brandenburg et al.(1995)]{bra95} Brandenburg, A., Nordlund, \AA., Stein, R. F., Torkelsson, U. 1995, 
      ApJ, 446, 741

\bibitem[Burbidge, Burbidge \& Prendergast(1963)]{bur63} Burbidge, E. M., Burbidge, G. R., Prendergast, K. H. 1963, 
      ApJ, 138, 375

\bibitem[Caproni, Abraham \& Mosquera Cuesta(2004)]{cap04} Caproni, A., Abraham, Z. \& Mosquera Cuesta, H. J. 2004, 
      ApJ, 616, L99

\bibitem[Caproni, Abraham \& Mosquera Cuesta(2006)]{cap06} Caproni, A., Abraham, Z., Mosquera Cuesta, H. J. 2006, 
      ApJ, 638, 120

\bibitem[Cecil, Wilson \& Tully(1992)]{cec92} Cecil, G., Wilson, A. S. \& Tully, R. B. 1992, 
      ApJ, 390, 365

\bibitem[Cecil et al.(2000)]{cec00} Cecil, G., Greenhill, L. J., DePree, C. G., Nagar, N., Wilson, A. S., 
      Dopita, M. A., P\'erez-Fournon, I., Argon, A. L., Moran, J. M. 2000, 
      ApJ, 536, 675

\bibitem[Court\`es \& Cruvellier(1961)]{cocr61} Court\`es, G., Cruvellier, P. 1961, 
      Compt. Rend. Acad. Sci. Paris, 253, 218

\bibitem[Elitzur(1992)]{elit92} Elitzur, M. 1992, 
      Astronomical Masers (Dordrecht: Kluwer).

\bibitem[Ford et al.(1986)]{for86} Ford, H. C., Dahari, O., Jacoby, G. H., Crane, P. C., Ciardullo, R. 1986,
      ApJ, 311, L7

\bibitem[Fragile \& Anninos(2005)]{fran05} Fragile, P. C. \& Anninos, P., 2005, 
      ApJ, 623, 347

\bibitem[Fruscione et al.(2005)]{fru05} Fruscione, A., Greenhill, L. J., Filippenko, A. V.,Moran, J. M., Herrnstein, J. R., Galle, E. 2005, 
      ApJ, 624, 103

\bibitem[Gammie, Narayan \& Blandford(1999)]{gam99} Gammie, C. F., Narayan, R., Blandford, R. 1999, 
      ApJ, 516, 177

\bibitem[Gammie et al.(2000)]{gam00} Gammie, C. F., Goodman, J., Ogilvie, G. I. 2000, 
      MNRAS, 318, 1005

\bibitem[Hawley, Gammie \& Balbus(1996)]{haw96} Hawley, J. F., Gammie, C. F., Balbus, S. A. 1996, 
      ApJ, 464, 690

\bibitem[Hawley \& Krolik(2001)]{hakr01} Hawley, J. F., Krolik, J. H. 2001, 
      ApJ, 548, 348

\bibitem[Herrnstein et al.(1997)]{her97} Herrnstein, J. R., Moran, J. M., Greenhill, L. J., Diamond, P. J., Miyoshi, M., 
      Nakai, N., Inoue, M. 1997, ApJ, 475, L17

\bibitem[Herrnstein et al.(1999)]{her99} Herrnstein, J. R., Moran, J. M., Greenhill, L. J., Diamond, P. J., Inoue, M., 
      Nakai, N., Miyoshi, M., Henkel, C., Riess, A. 1999, 
      Nature, 400, 539

\bibitem[Herrnstein et al.(2005)]{her05} Herrnstein, J. R., Moran, J. M., Greenhill, L. J., Trotter A. S. 2005, 
      ApJ, 629, 719.

\bibitem[Ho et al.(1997)]{ho97} Ho, L. C., Filippenko, A. V., Sargent,W. L.W., Peng, C. Y. 1997, 
      ApJS, 112, 391

\bibitem[Ivanov \& Illarianov(1997)]{ivil97} Ivanov, P. B., Illarianov, A. F. 1997, 
      MNRAS, 285, 394
	  
\bibitem[King, Pringle \& Livio(2007)]{kin07} King, A. R., Pringle, J. E., Livio, M. 2007, 
      MNRAS, 376, 1740

\bibitem[Kumar \& Pringle(1985)]{kupr85} Kumar, S., Pringle, J. E. 1985, 
      MNRAS, 213, 435

\bibitem[Larwood(1997)]{larw97} Larwood, J. D. 1997, 
      MNRAS, 290, 490

\bibitem[Lasota et al.(1996)]{las96} Lasota, J. P., Abramowicz, M. A., Chen, X., Krolik, J., Narayan, R., Yi, I. 1996, 
      ApJ, 462, 142

\bibitem[Lodato \& Pringle(2006)]{lopr06} Lodato, G., Pringle, J. E. 2006, 
      MNRAS, 368, 1196

\bibitem[Lubow, Ogilvie \& Pringle(2002)]{lub02} Lubow, S. H., Ogilvie, G. I. \& Pringle, J. E. 2002, 
      MNRAS, 337, 706

\bibitem[Maloney, Begelman \& Pringle(1996)]{mal96} Maloney, P. R., Begelman, M. C., Pringle, J. E. 1996, 
      ApJ, 472, 582

\bibitem[Maloney, Begelman \& Nowak(1998)]{mal98} Maloney, P. R., Begelman, M. C., Nowak, M. A. 1998, 
      ApJ, 504, 77

\bibitem[Markovi\'c \& Lamb(1998)]{mala98} Markovi\'c, D., Lamb, F. K. 1998, 
      ApJ, 507, 316

\bibitem[Martin et al.(1989)]{mar89} Martin, P., Roy, J.-R., Noreau, L., Lo, K. Y. 1989, 
      ApJ, 345, 707

\bibitem[Menou et al.(2000)]{men00} Menou, K., Hameury, J.-M., Lasota, J.-P., Narayan, R. 2000, 
      MNRAS, 314, 498

\bibitem[Miyoshi et al.(1995)]{miy95} Miyoshi, M., Moran, J. M., Herrnstein, J. R., Greenhill, L. J., Nakai, N.,
      Diamond, P. J., Inoue, M. 1995, Nature, 373, 127

\bibitem[Modjaz et al.(2005)]{mod05} Modjaz, M., Moran, J. M., Kondratko, P., Greenhill, L. J. 2005, 
      ApJ, 626, 104

\bibitem[Moran, Greenhill \& Herrnstein(1999)]{mor99} Moran, J. M., Greenhill, L. J., Herrnstein, J. R. 1999, 
      J. Astrophys. Astron., 20, 165

\bibitem[Natarajan \& Pringle(1998)]{arpr98} Natarajan, P., Pringle, J. E. 1998, 
      ApJ, 506, L97

\bibitem[Natarajan \& Armitage(1999)]{naar99} Natarajan, P., Armitage, P. J. 1999, 
      MNRAS, 309, 961

\bibitem[Nelson \& Papaloizou(2000)]{nepa00} Nelson, R. P., Papaloizou, J. C. B. 2000, 
      MNRAS, 315, 570

\bibitem[Neufeld \& Maloney(1995)]{nema95} Neufeld, D. A., Maloney, P. R. 1995, 
      ApJ, 447, L17

\bibitem[Ogilvie(1999)]{ogil99} Ogilvie, G. I. 1999, 
      MNRAS, 304, 557

\bibitem[Papaloizou \& Pringle(1983)]{papr83} Papaloizou, J. C. B., Pringle, J. E. 1983, 
      MNRAS, 202, 1181

\bibitem[Papaloizou \& Lin(1995)]{pali95} Papaloizou, J. C. B., Lin, D. N. C. 1995, 
      ApJ, 438, 841

\bibitem[Papaloizou \& Terquem(1995)]{pate95} Papaloizou, J. C. B., Terquem, C., 1995, 
      MNRAS, 274, 987

\bibitem[Papaloizou, Terquem \& Lin(1998)]{pap98} Papaloizou, J. C. B., Terquem, C., Lin, D. N. C. 1998, 
      MNRAS, 497, 212

\bibitem[Pringle(1996)]{prin96} Pringle, J. E. 1996, 
      MNRAS, 281, 857

\bibitem[Rees(1978)]{rees78} Rees, M. J. 1978, 
      Nature, 275, 516

\bibitem[Sakimoto \& Coroniti(1981)]{saco81} Sakimoto, P.J. \& Coroniti, F., 1981, 
      ApJ, 247, 19

\bibitem[Scheuer \& Feiler(1996)]{scfe96} Scheuer P. A. G., Feiler R. 1996, 
      MNRAS, 282, 291

\bibitem[Shakura \& Sunyaev(1973)]{shsu73} Shakura, N. I., Sunyaev, R. A. 1973, 
      A\&A, 24, 337

\bibitem[Shapiro \& Teukolsky(1983)]{shte83} Shapiro S. L. \& Teukolsky S. A., 1983, 
      In: Black Holes, White Dwarfs, and Neutron Stars. John Wiley \& Sons, NY, p.362

\bibitem[Siemiginowska \& Czerny(1989)]{sicz89} Siemiginowska, A., Czerny, B. 1989, 
      MNRAS, 239, 289

\bibitem[Smak(1999)]{smak99} Smak, J. 1999, 
      Acta Astron. 49, 391

\bibitem[Stella \& Vietri(1998)]{stvi98} Stella, L., Vietri, M. 1998, 
      ApJ, 492, L59

\bibitem[Stone et al.(1996)]{sto96} Stone, J. M., Hawley, J. F., Gammie, C. F., Balbus, S. A. 1996, 
      ApJ, 463, 656

\bibitem[van Albada \& van der Hulst(1982)]{alhu82} van Albada, G. D., \& van der Hulst, J. M. 1982, 
      A\&A, 115, 263

\bibitem[van der Kruit, Oort, \& Mathewson(1972)]{kru72} van der Kruit, P. C., Oort, J. H., Mathewson, D. S. 1972, 
      A\&A, 21, 169

\bibitem[van der Kruit(1974)]{kru74} van der Kruit, P. C. 1974, 
     ApJ, 192, 1

\bibitem[Wilkins(1972)]{wilk72} Wilkins, D. C. 1972, 
      PRD, 5, 814

\bibitem[Wilson, Yang \& Cecil(2001)]{wil01} Wilson, A. S., Yang, Y., Cecil, G. 2001, 
      ApJ, 560, 689

\bibitem[Winters, Balbus \& Hawley(2003)]{win03} Winters, W. F., Balbus, S. A., Hawley, J. F. 2003, 
      ApJ, 589, 543

\end{thebibliography}
\end{document}